# Spatial sampling design to improve the efficiency of the estimation of the critical parameters of the SARS-CoV-2 epidemic

Alleva, G., Arbia, G., Falorsi, P.D., Nardelli, V., Zuliani, A.

**Abstract:** The pandemic linked to COVID-19 infection represents an unprecedented clinical and healthcare challenge for many medical researchers attempting to prevent its worldwide spread. This pandemic also represents a major challenge for statisticians involved in quantifying the phenomenon and in offering timely tools for the monitoring and surveillance of critical pandemic parameters. In a recent paper, Alleva *et al.* (2020) proposed a two-stage sample design to build a continuous-time surveillance system designed to correctly quantify the number of infected people through an indirect sampling mechanism that could be repeated in several waves over time to capture different target variables in the different stages of epidemic development. The proposed method exploits the indirect sampling (Lavalle, 2007; Kiesl, 2016) method employed in the estimation of rare and elusive populations (Borchers, 2009; Lavallée and Rivest, 2012) and a capture/recapture mechanism (Sudman, 1988; Thompson and Seber, 1996). In this paper, we extend the proposal of Alleva *et al.* (2020) to include a spatial sampling mechanism (Müller, 1998; Grafström *et al.*, 2012, Jauslin and Tillè, 2020) in the process of data collection to achieve the same level of precision with fewer sample units, thereby facilitating the process of data collection in a situation where timeliness and costs are crucial elements. We present the basic idea of the new sample design, analytically prove the theoretical properties of the associated estimators and show the relative advantages through a systematic simulation study where all the typical elements of an epidemic are accounted for.

1. Introduction

The SARS-CoV-2 pandemic has affected in Western countries in a sudden and devastating way. In most cases, the data describing its evolution were generally collected without an organic design. Those related to hospital admissions and derived from the testing processes that were gradually activated (i.e., number of diagnostic tests carried out and their results) were organized. Information on gender and age was generally collected.
To tackle the pandemic, non-pharmaceutical interventions (NPIs) (such as lockdown and social distancing) have been successfully adopted and, in most cases, generalized for the entire population. This approach reduced the need to have a detailed information framework for a territory, at least during the first wave.
However, the subsequent waves of the pandemic and the presence of numerous mutations of the virus, together with the possibility of future new pandemic emergencies, require the establishment of relevant and timely monitoring and surveillance tools to control some of the critical parameters by adding to the existing health surveillance data sample data collected *ad hoc*.

Indeed, sample surveys used to estimate the prevalence of the epidemic are of paramount importance since they allow the detection of asymptomatic and paucisymptomatic cases that are not observed with officially collected medical swabs, which are mainly directed toward symptomatic cases (Aguilar *et al*., 2020; Chugthai *et al*., 2020; Li *et al*., 2020; Mizumoto *et al*., 2020a, 2020b and Yelin *et al*., 2020) and, sometimes, their contacts, with an obvious underestimation of prevalence and an overestimation of the lethality rate. The risk of erroneous inference based on such data has been stigmatized (Ioannidis, 2020). To identify possible risk factors, surveys should also be extended to observe the behaviour of the sample units in the previous two or three weeks, as these units are relevant to understanding the spread dynamic of the infection, as well as some of their demographic and social characteristics. The infected people who emerge from the sample survey can form a panel on which to estimate the changes in status over time.

However, designing a sample survey that is not only reliable but also realistic and efficient in its implementation constitutes a major challenge for the statistical community. In a recent paper (Alleva *et al*., 2020), we proposed a two-stage sampling design for building a continuous-time surveillance system to assess the prevalence of infected people in the population. The design is defined considering an indirect sampling mechanism, and it is suitable to consider various target variables during different stages of the epidemic. The statistical optimality properties of the proposed estimators were formally proven and tested with a Monte Carlo experiment: the estimators are unbiased and are much more efficient than the simple random sampling scheme.

Although replicable independently at different geographical levels, the quoted methodology does not contain any reference to the spatial correlation, which, in contrast, represents an intrinsic feature of many empirical phenomena and, in particular, infectious diseases (Cliff *et al*., 1981). In this paper, we extend the proposal of Alleva *et al*. (2020) to include a spatial sampling mechanism in the process of data collection to achieve the same level of precision with fewer sample units, thus helping the process of data collection in a situation where timeliness is a crucial element. We present the basic idea of the new sample design, the theoretical properties of the associated estimators and the relative advantages through a systematic simulation study where the main typical elements of an epidemic are accounted for. Indeed, to properly evaluate the advantages of introducing spatial elements in the data collection mechanism, many aspects need to be considered, both in the sampling scheme and in the spatial mechanism of infection diffusion. In our model, we consider a multiplicity of these aspects, and we present the results of simulations to evaluate their relative contribution to the accuracy of the proposed method.

In summary, the proposed sampling scheme is a general two-stage mechanism with a first stage constituted by a sample of *clusters (areas)* selected with inclusion probabilities proportional to their size (PPSs) and a second stage where the units are selected from the sampled cluster via a simple random sampling procedure. In the first stage, the clusters are drawn by following an explicit spatial sampling design. In particular, we consider the cube algorithm (Chauvet and Tillé 2006) with different balancing elements and spatial sampling (Tillé, 2020).

The remainder of the paper is organized as follows. In Section 2, the main results of Alleva *et al*. (2020) together with a literature review of the most popular spatial sampling designs are summarized. The basic sampling framework of the new method proposed together with the specific first-stage sampling design are described in Sections 3 and 4. The results of a Monte Carlo study of the proposed method are presented and discussed in Section 5. Conclusions and future challenges are highlighted in Section 6.

## 2. Previous proposals for monitoring COVID-19 and the importance of spatial sampling

Starting from the observation that official medical swabs are directed mainly toward infected people and therefore largely underestimate the number of infected people, Alleva *et al.* (2020) proposed a sampling strategy to estimate the total infected population, including those that show no symptoms, to better estimate mortality and lethality rates. The proposed procedure preliminarily involves dividing the population into two groups. The first group (Group A) consists of all the individuals who have a verified state of infection together with those who had contact with them. The second group (Group B), in contrast, contains both healthy people and those who are still in a phase of incubation with no symptoms. Obviously, the proportion of infected people is much larger in the first group than in the second group, so focusing on their contacts is a way of maximizing the number of infected people included in the sample. However, including the second group is also necessary to correctly estimate the epidemic parameters. For both groups, Alleva *et al.* (2020) proposed two distinct methodologies based on the idea of *indirect sampling* (Lavalle, 2007; Kiesl, 2016). In particular, a sample from Group A is partly drawn without replacement from the whole infected population and partly drawn without replacement from all their contacts. Furthermore, the design also included a traditional population panel survey with sample rotation to inspect group B and was associated with an indirect sampling mechanism to trace and sample the individuals who came in contact with the infected people. The competitive advantage of the proposal of Alleva *et al.* (2020) with respect to other sampling plans (e.g., those based exclusively on indirect sampling or only on the panel sample) relies mainly on the combination of the two sampling strategies, which produce a flexible tool designed to closely monitor the dynamics of the epidemics in their different phases. In the quoted paper, the authors provided formal proof of the unbiasedness of their proposed design and of its greater efficiency with respect to simple random sampling and reinforced their conclusions with a set of simulations.

The present work builds upon the contribution of Alleva *et al.* (2020) and extends it by considering a spatial sample design, both in the first and second groups. Our aim is to further increase the estimator's efficiency to achieve the same level of precision with fewer sample units, thus reducing the number of tests and their associated costs, which is an important concern in the epidemic monitoring and surveillance phase.

In the statistical literature, *spatial sampling* is the process through which observations are collected in a two-dimensional framework with specific attention to their location (Müller, 2007; Wang *et al.*, 2012) to maximize the probability of capturing the spatial variability of the empirical phenomena. Indeed, geographically distributed data are usually characterized by significant spatial correlation, which is especially true if they refer to contagious diseases (Cliff *et al.*, 1981). Spatial sampling is therefore distinct from conventional sampling, where data are assumed to be independent and identically distributed. A rather popular spatial sampling strategy, initiated by Arbia (1993), consists of exploiting spatial correlation to maximize the information content while minimizing the sample size, thus reducing the overall costs. This strategy was termed DUST after the acronym of the dependent areal units sequential technique, and it was inspired by model-based assumptions on the dependence of the random field generating the data. The method was characterized by variable inclusion probabilities at each step (Brewer and Hanif, 1983) and extended the idea of the balanced sampling design excluding continuous units (BSEC) introduced by Hedeyat *et al.* (1988) for unidimensional data. The DUST method consists of a draw-by-draw scheme: starting from a unit selected at random, it suggests updating the selection probabilities by considering, at each step, the spatial correlation and the distance with the units already in the sample. A consequence of the procedure is that the study area is not sampled uniformly: the higher the (positive) spatial correlation is, the further away we observe the next sample unit in the sequential collection process because part of the information content is already contained in the units already sampled. Many more examples are found in the extensive literature that followed based on the use of contiguous units (see, among others, Hedayat, Rao, and Stufken, 1988; Arbia and Lafratta, 2002, Stevens and Olsen, 2004; Arbia *et al.*, 2007, Mandal, Parsad, Gupta, and Sud, 2009; Wright and Stufken, 2008; Tillè *et al.*, 2018; Fattorini *et al.*, 2020). More recently, some new sampling methodologies that explicitly exploit the interpoint distances between locations in the selection procedure, such as local pivotal (LP) methods (Grafström *et al.*, 2012), which were derived as an extension of the pivotal methods introduced by Deville and Tille (1998), have been proposed. In particular, similar to Arbia (1993), Grafström *et al.* (2012) suggested employing an updating rule of the probabilities of inclusion that, at each step, locally keeps the sum of the probabilities as constant as possible. Further alternatives are the local cube method (LC, Deville and Tillé, 2004), which allows selection probabilities to be balanced on several auxiliary variables, and the spatially correlated Poisson sampling method proposed by Grafström (2012).

In this paper, we expand the results of Alleva *et al.* (2020) by introducing local pivotal and local cube methods when sampling both population groups in the original design to produce more parsimonious sample plans in terms of the sample size. This is particularly desirable to reduce the costs and time involved in data collection during periods of pandemic emergency, thus facilitating monitoring, control and surveillance of infection diffusion.

## 3. The basic sampling framework

### 3.1. Population and target parameter

Before presenting our proposed sample design, let us start by defining $U$ as the population of interest of size $N$ and suppose that it can be partitioned into $M$ subpopulations, called clusters (or areas) denoted as $U_1, \ldots, U_i, \ldots, U_M$. The set of clusters is symbolically represented as $U_I = \{1, \ldots, i, \ldots, M\}$. Cluster $U_i$ has $N_i$ units, with

$$N = \sum_{i=1}^{M} N_i.$$

Let $v_{ij}$ ($i = 1, \ldots, M$; $j = 1, \ldots, N_i$) be a dichotomous variable $v$ that assumes a value of 1 if person $j$ of cluster $U_i$ has a verified state of infection and a value of 0 otherwise. $v_{ij} = 1$ indicates that person $ij$ could be either hospitalized or in compulsory quarantine. Let

(1) $\quad \mathcal{V}_i = \sum_{j=1}^{N_i} v_{ij} \quad and \quad \mathcal{V} = \sum_{j=1}^{M} \mathcal{V}_i$

be the known totals of the verified infected people in cluster $U_i$ and in population $U$. The quantities $\mathcal{V}_i$ and $\mathcal{V}$ are generally known from informative health care systems.

Let $y_{ij}$ be the value of variable $y$ for person $ij$, where $y$ is equal to 1 if the person is infected and 0 otherwise. If $v_{ij} = 1$, then obviously $y_{ij} = 1$; however, if $v_{ij} = 0$, then it is possible that either $y_{ij} = 1$ (an infected person for whom the infection has not yet been verified) or $y_{ij} = 0$ (a healthy person).

The target parameter, $Y$, is the total number of infected people (verified or not), that is

$$(2)\ Y = \sum_{i=1}^{m} \sum_{j=1}^{N_i} y_{ij} = \sum_{j=1}^{N_i} Y_i,$$

where $Y_i$ indicates the number of infected people of cluster $i$.

### 3.2. The general sampling scheme

We select sample $S$ by a general two-stage sampling design without replacement.

A first-stage sample, $S_I$, of fixed size $m$ is selected without replacement from $U_I$, with inclusion probabilities $\pi_{Ii}$ $(i = 1,2 \dots, M)$.

A standard solution is to select cluster $i$ with a probability proportional to its size (PPS):

$$(3)\ \pi_{Ii} = m \frac{N_i}{N}.$$

In the second stage of sampling, a second stage sample, $S_{IIi}$, of fixed size $\bar{n}$ is selected from sample cluster $i$ by *drawing the units* without replacement via a simple random sampling procedure. The second-stage inclusion probability $\pi_{IIi}$ of people in the sampled cluster $i$ is

$$(4)\ \pi_{IIi} = \frac{\bar{n}}{N_i}.$$

The final inclusion probability of person $j$ being selected from cluster $i$ is

$$(5)\ \pi_{ij} = \pi_{Ii} \pi_{IIi}$$

If the first-stage inclusion probabilities, $\pi_{Ii}$, are defined according to Formula 3, the sampling process is *self-weighting* (Murthy and Sethy, 1965) in the sense that all the units in $U$ have an equal probability of being selected, irrespective of their cluster. Indeed,

$$(6)\ \pi_{ij} = m \frac{N_i}{N} \frac{\bar{n}}{N_i} = m \frac{\bar{n}}{M}.$$

The *self-weighting* property defines a sampling design that is somehow optimal (Kish, 1966) in the sense that it avoids the negative impact of the variability of the sampling weights on the sampling variances.

We test the status of infection on each of the $m \times \bar{n}$ people selected in the sample, thus knowing the value of variable $y_{ij}$.

The Horvitz Thompson (Narain, 1951; Horvitz Thompson, 1952) estimator of $Y$ is

$$(7)\ \hat{Y} = \sum_{i=1}^{m} \sum_{j=1}^{\bar{n}} y_{ij} w_{ij} = \sum_{i=1}^{m} \hat{Y}_i w_{Ii},$$

where

(8) $w_{ij} = w_{Ii} w_{IIi} = \dfrac{1}{\pi_{ij}}$ and $\hat{Y}_i = \sum_{j=1}^{n_i} y_{ij} w_{IIi}$,

with

$w_{Ii} = \dfrac{1}{\pi_{Ii}}$ and $w_{IIi} = \dfrac{1}{\pi_{IIi}}$

being the first-stage and the second-stage sampling weights, respectively.

### 3.3. The anticipated variance

The spatial correlation of the units may be formalized by the following working model (WM) *M*, according to which

(9) $y_{ij} = m(\boldsymbol{x}_{ij}; \boldsymbol{\beta}) + u_{ij}$,

where $\boldsymbol{x}_{ij}$ is a column vector of auxiliary variables specific for unity $ij$, $m(\boldsymbol{x}_{ij}; \boldsymbol{\beta})$ is a known function applied to the column vector of auxiliary variable $\boldsymbol{x}_{ij}$, $u_i$ is a random residual, and $\boldsymbol{\beta}$ is the unknown column vector of the model parameters. Let $E_M(\cdot), V_M(\cdot),$ and $Cov_M(y_{ij}, y_{\ell k})$ denote the model expectation, variance and covariance, respectively. The spatial correlation of the units may be formalized as

(10) $E_M(y_{ij}) = m(\boldsymbol{x}_{ij}; \boldsymbol{\beta}), V_M(y_{ij}) = \sigma_u^2, Cov_M(y_{ij}, y_{\ell k}) = \sigma_u^2 \rho_{ij,\ell k}$,

where $\rho_{ij,\ell k}$ is the model correlation between units $ij$ and $\ell k$, and $\sigma_u^2$ is a variance scalar factor. $\rho_{ij,\ell k}$ is defined as

(11) $\rho_{ij,\ell k} = \rho[d(ij, \ell k)]$

where $|\rho(\cdot)| \leq 1$ is a decreasing function of the distance, $d(ij, \ell k)$, between units $ij$ and $\ell k$. For instance, Grafström and Tillé (2013) propose to specify this function as

$$\rho[d(ij, \ell k)] = \rho^{d(ij,\ell k)}$$

where $-1 \leq \rho \leq 1$. Shabenberger and Gotway (2005) and Banerjee *et al.* (2004) suggest adopting an isotropic Gaussian covariance function defined as

$\rho_{ij,\ell k} = exp\{-3[(d(ij,\ell k))/\alpha]^2\} + \tau^2;$ if $d(ij,\ell k) \geq 0$, 0 otherwise

where $\tau^2$ is the nugget effect and $\alpha$ is the empirical range, that is, the distance at which correlation decreases to less than 0.05. The inverse of $\alpha$, $\phi$, is known as the decay parameter that controls the rapidity with which the covariance declines with increasing distance. The normalization factor of 3 is not essential but is rather common in geostatistics.

Using model (9), with model expectations given by formulae (10) and (11), the anticipated variance of $\hat{Y}$ is (see Appendix 1)

$$(12) \quad E_P E_M (\hat{Y} - Y)^2 = \sum_{i=1}^{M} \frac{1}{\pi_{Ii}} \sigma_i^2 - F,$$

where $E_P(\cdot)$ denotes the expectation over repeated sampling, with

$$(13) \quad \sigma_i^2 = \eta_i^2 + \sigma_{\bar{y}_i}^2 N_i \left( \frac{N_i - \bar{n}}{\bar{n}} \right) + N_i^2 \frac{\sigma_u^2}{\bar{n}} + \rho_{i,i} \sigma_u^2 + \sum_{\ell \neq i}^{M} \frac{\pi_{Ii,I\ell}}{\pi_{I\ell}} \sigma_u^2 \rho_{i,\ell} \text{ and}$$

$$(14) \quad F = \sum_{i=1}^{M} \left\{ \eta_i^2 + N_i^2 \sigma_u^2 + \rho_{i,i} \sigma_u^2 + \sum_{\ell \neq i}^{M} \sigma_u^2 \rho_{i,\ell} \right\}.$$

$$(15) \quad \rho_{i,\ell} = \sum_{j=1}^{N_i} \sum_{k=1}^{N_\ell} \sigma_u^2 \rho[d(ij, \ell k)], \quad \text{and } \rho_{i,i} = \sum_{j=1}^{N_i} \sum_{k \neq j}^{N_i} \sigma_u^2 \rho[d(ij, ik)],$$

in which $\eta_i^2$ and $\sigma_{\bar{y}_i}^2$ are defined in Formulae A3a and A5 of Appendix A1, and $\pi_{Ii,I\ell}$ is the joint inclusion probability of selecting clusters $i$ and $\ell$ in first-stage sampling.

**Remark 1. A note on the feasibility of the proposed sampling design.** The proposed sampling design has good feasibility in practical contexts. For instance, in Italy, each area (or cluster) could coincide with a local health unit's territory, which is the administrative zone in which public health is managed. Each local area could be easily made autonomous in drawing an SRS design in its territory. The uniformity of the second-stage sampling sizes makes it possible to define general guidelines to lead local-level field operations. This could be straightforwardly managed at the local level. Moreover, costs can be determined very simply by multiplying the unit cost for sampling $m$ first-stage units by the unit cost for sampling $\bar{n}$ people in each sampled cluster.

## 4. Specification of the first-stage sampling design

### 4.1 Generalities

Clusters can be sampled with different algorithms, thus leading to specific first-stage sampling designs. The basic techniques considered here are the cube algorithm (Chauvet and Tillé 2006) and spatial sampling (Tillé, 2020). The cube algorithm ensures that the Horvitz-Thompson (HT) estimates of the selected first-stage sample reproduce (at least approximately) the known characteristics of some auxiliary variables available for population $U_I$. This can be expressed as

$$(16) \quad \sum_{i \in S_I} \frac{\boldsymbol{d}_{Ii}}{\pi_{Ii}} \cong \sum_{i \in U_I} \boldsymbol{d}_{Ii},$$

where $\boldsymbol{d}_{Ii}$ is a vector of $H$ auxiliary variables for cluster $i$.

Spatial sampling avoids the joint selection of units that are positively correlated. As the correlated units are generally geographically close, the sample must be well spread (or spatially balanced) in the territory.

The first-stage sampling designs we consider are described below.

*i)    The fixed-size probability proportional to size (FPPS) sampling design*

We can implement this design by the cube algorithm, collapsing balancing variables $\mathbf{d}_{Ii}$ to scalar value $\pi_{Ii}$. Therefore, the balancing equations (16) are defined as

(17) $$\sum_{i \in S_I} \frac{\pi_{Ii}}{\pi_{Ii}} = \sum_{i \in U_I} \pi_{Ii} = m.$$

*ii)    The cube method based on the verified infected people(CBV)*

This method ensures that the first-stage sampling size is fixed and that first-stage HT estimates of the total number of verified infected individuals reproduce the known totals.

The balancing variables are $\mathbf{d}_{Ii} = (\pi_{Ii}, \mathcal{V}_i)'$, and the balancing equations become

(18) $$\sum_{i \in S_I} \frac{1}{\pi_{Ii}} (\pi_{Ii}, \mathcal{V}_i)' = (m, \mathcal{V}).$$

*iii)    The local pivotal (LP) method*

Grafström (2012) introduced the local pivotal method, which enables the selection of unequal probability samples that are well spread over the population. This method uses the distance between units to create small joint inclusion probabilities for nearby units, forcing the samples to be well dispersed. The sample is fixed in size, and the first-stage inclusion probabilities are equal to the planned $\pi_{Ii}$ ($i = 1, \dots, M$).

*iv)    The local cube method based on verified infected (LCBV)*

The local cube method (Grafström and Tillé, 2013) is an extension of the LP method, which selects a sample that is both spread out geographically and balanced in terms of auxiliary variables. The algorithm consists of running the flight phase of the cube method on a subset of $H + 1$ neighbouring units, with $H$ being the number of balancing equations on which we want to balance sampling. The balancing equations of the LCBV methods are the same as those of the CBV method.

*v)    The local cube method based on the geography of the phenomenon (LCBG)*

In this case, the balancing variables are given by a transformation of the geographical coordinates of the cluster. Therefore, the sample is well spread and reproduces the known spatial distribution of $U_I$.

**Remark 2. A note on the efficiency of the proposed sampling design.** Looking at expressions (12), (13), and (15), we can see that we may greatly increase the efficiency of the strategy by geographically dispersing and balancing the sample in both stages of the design. As demonstrated in Appendix A2, using model (9), a very efficient two-stage sampling design consists of the following:

1. using a PPS first-stage sampling design balanced on variables $\mathbf{d}_{Ii,aug} = \left(\mathbf{d}'_{Ii}, \mathbf{x}'_{Ii}\right)'$, where $\mathbf{x}_{Ii} = \sum_{j=1}^{N_i} \mathbf{x}_{ij}$ is the total number of $\mathbf{x}_{ij}$ in cluster $i$;
2. selecting a well-spread sample (or spatially balanced) of clusters;
3. drawing a second-stage sampling design balanced on independent variables $\mathbf{x}_{ij}$; and
4. selecting a well-spread second-stage sample of units.

Adopting the above strategy, the optimal first-stage inclusion probabilities are the PPSs given in formula 3.

**Remark 3. A note on the role of the model.** The HT estimator will be efficient if the population is close to the realization of model (9) but maintains desirable properties, such as design unbiasedness and consistency, irrespective of the shape of the finite population scatter. It follows that the estimator is design-unbiased, irrespective of whether the assumptions of the model are true or false. On the other hand, the appropriateness of the model is crucial to achieving a small variance.

## 5. A Monte Carlo evaluation of the proposed method

### 5.1. Simulation design

Given the lack of official data at the individual level and the limited ability of health systems to track all infections during the pandemic, we test the proposed sampling methodologies on a generated dataset representing an artificial population. Each individual of the population can be classified as susceptible, infected or removed according to the framework originally proposed by Hamer (1906), Kermack and McKendrick (1927) and Soper (1929), known as the "SIR model". Among the numerous variations to the framework, Alleva *et al.* (2020) adapted the SIR model to better represent the characteristics of the SARS-CoV-2 epidemic, generating an artificial population with six categories of individuals, namely, susceptible people (S), those exposed to the virus (E), those infected with symptoms (I), those without symptoms (A) and those that are removed from the population either because they healed (R) or are dead (D). In this experiment, we considered all the subjects belonging to group (I) to be "known" infected (identified by a positive swab from health screening), while in group A, we considered all the "unknown" infected subjects, i.e., all those who were not aware of being infected. This division is crucial to correctly model the transmission chain. All infected individuals who are not aware of being infected (A) are not isolated in quarantine and continue to move and meet other people, with a high risk of infecting them.

In our simulation study, we test our proposed sampling methodology on a randomly generated dataset representing an artificial population.

The general structure of the simulation can be described as follows.

Each individual of the population is classified into the six previously mentioned categories. The random generation model is parameterized with respect to the time spent in each category and the probability of passing from one state to another (transition matrix).

For the map generation process, we considered a population distributed across 400 spatial units laid on a regular 20-by-20 square lattice grid. The density of the population residing in each cell, which is equivalent to the susceptible population at day 0, was generated considering different spatial distributions.

The movements of the individuals were simulated considering that, on each simulation day, one subset of the population moves towards the centre of the map, while another subset of the population moves uniformly by one cell. After these movements, social interactions take place in each cell, and all individuals return to the cell from which they originated (i.e., residence). Every day, contagion is simulated to happen during the meeting of individuals in the various movements. The number of meetings in each cell and the number of people involved in the movements are also randomly determined.

Epidemic curves are simulated with the mechanism of mobility and social interaction divided into two phases: in the first four weeks (Phase 1), the interaction corresponds to a situation of normality, while in the following six weeks (Phase 2), a state of lockdown was simulated.

To measure the impact of the spatial correlation of the susceptible population and mobility during the various phases, we calculated the Moran index for each day of the known (I) and total infected (I+A) populations. The impact on the spatial correlation of the infected due to the mechanisms of mobility and social interactions is far greater than that induced by the geographical distribution of the population.

Figure 1: The six basic categories of our simulation model and their transition patterns (From Alleva *et al.* 2021).

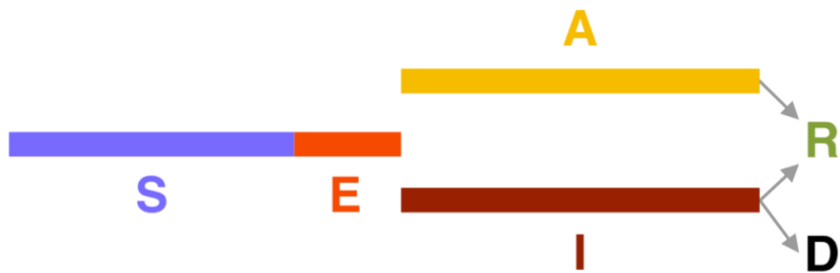

Going into more detail, the random generation process is fundamentally parameterized by the time spent in each category and by the probability of passing from one state to another. In particular, following evidence of COVID-19 diffusion, we assume that each individual remains in the exposed state (E) for 5 days after infection, while the state of infection with symptoms (I) and without symptoms (A) has a maximum duration of 14 days. Regarding the probabilities of switching between categories, we assume that an individual in the exposed state (E) has a probability of 0.25 of passing to the infected with symptoms (I) state and a probability of 0.75 of becoming infected without symptoms (A). After the infection period, the probability of dying for an individual with symptoms (I) is 0.15 (death rate case), while all the other infected individuals pass to the removed state (R). The full transition matrix is reported in Table 1. The probability of an individual being infected (i.e., the probability of passing from a susceptible state (S) to exposed (E) state) is not explicitly derived but depends on mobility and social interactions. All the chosen values can obviously be parametrized differently, but our choices seem realistic in view of reproducing the dynamics of the SARS-CoV-2 pandemic, as manifested in the various waves in 2020 and 2021.

Table 1: Transition matrix.

|   | E | I | A | R | D |
|---|---|---|---|---|---|
| E | 0 | 0.25 | 0.75 | 0 | 0 |
| I | 0 | 0 | 0 | 0.85 | 0.15 |
| A | 0 | 0 | 0 | 1 | 0 |
| R | 0 | 0 | 0 | 1 | 0 |
| D | 0 | 0 | 0 | 0 | 1 |

*Note: the probability of passing from state S to state E depends on mobility and social interactions.*

For the map generation process, we considered a population distributed across 400 spatial units laid on a regular 20-by-20 square lattice grid. The structure of the map is intentionally generic: in fact, it can represent both a city divided into neighbourhoods and a small region divided into several cities. The density of the population residing in each cell, which is equivalent to the susceptible population at day 0, was generated considering different spatial distributions. In particular, we generated 20,000 individuals distributed with a spatial autocorrelation parameter ($\rho$) equal to 0.3, 0.5 or 0.7 (as reported in Figure 2) to reproduce different patterns of spatial agglomeration in urban situations (Xu *et al.*, 2010). The experiment was also carried out with very high spatial autocorrelation (0.9) and, alternatively, with no spatial autocorrelation, but these results are omitted for brevity because they represent extreme cases very rarely found in practical instances.

**Figure 2: Generated grid map of a simulated population with different levels of spatial correlation.**

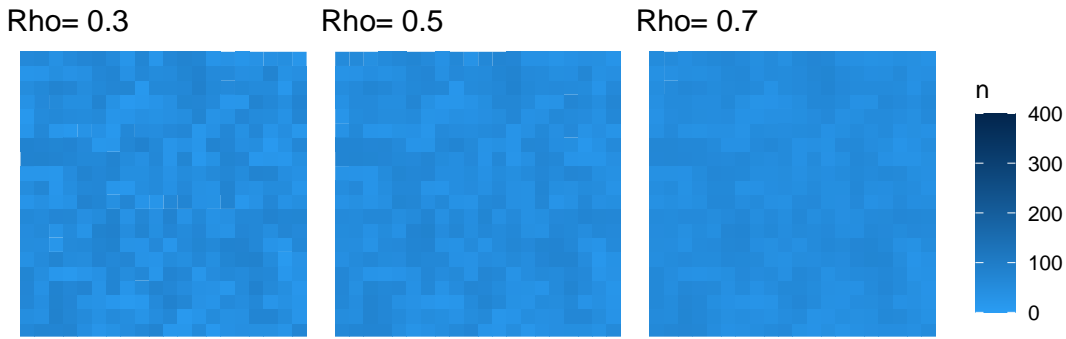

The movements of the individuals were simulated as follows: for each simulation day, one subset of the population ($m_1$) moves towards the centre of the map, while another subset ($m_2$) of the population moves uniformly by one cell. After these movements, social interactions take place in each cell, and all individuals return to the cell from which they originated (i.e., residence). We built this simulation by considering the movements of individuals who go daily to the city centre or central business district (CBD) (the four central cells of the map are considered the points of attraction) for work or leisure. On a higher geographical level, the map could also represent a small region where the inhabitants of the smaller cities frequently commute to the larger towns.

Every day, contagion is simulated to happen during the meeting of individuals and during their various movements in the geographical space. A number $i_m$ of susceptible people are considered newly infected people (and they are moved then into *exposed state* E) if at least one asymptomatic or exposed person is present in the meeting. The number of meetings in each cell is determined by a random number drawn from a Poisson distribution with a parameter, say, $c_n$, while the number of people involved in the movements is also a Poisson number but characterised by a different parameter, $c_p$.

Epidemic curves are simulated with the mobility and social interaction mechanism and divided into two phases: in the first four weeks (Phase 1), the interaction corresponds to a situation of normality, while in the following six weeks (Phase 2), a state of lockdown was simulated. The parameters for each phase are reported in Table 2.

Table 2: Simulation parameters for each phase.

|          | Phase 1  | Phase 2 |
|----------|----------|---------|
| Duration | 4 weeks  | 6 weeks |
| $m_1$    | 10%      | 1%      |
| $m_2$    | 5%       | 0%      |
| $c_n$    | 5        | 2       |
| $c_p$    | 5        | 3       |
| $i_m$    | 2        | 1       |

All the procedures described so far have led to the generation of various epidemic curves. The result of a single run of the simulation is reported in Figure 3 as an example. The trend of the various categories of the model roughly corresponds to what has been observed in reality in the different Italian regions affected by the SARS CoV-2 pandemic during 2020.

Figure 3: Epidemic curves for maps with different levels of autocorrelation

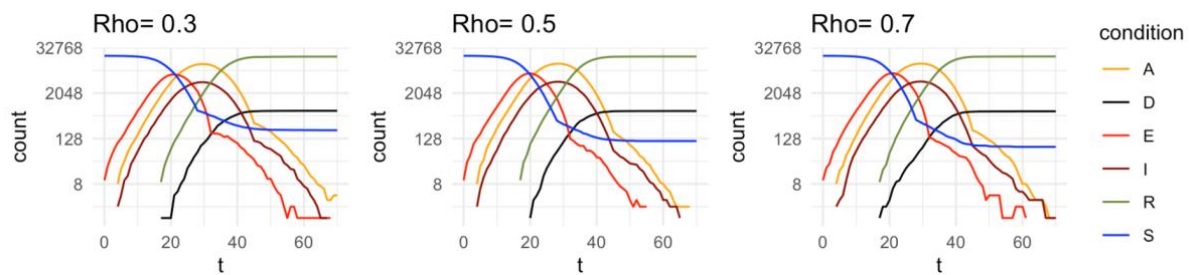

To measure the impact of the spatial correlation of the susceptible population and mobility during the various phases, we calculated the Moran index (Moran, 1950) for each day of both the known infected (I) and the total infected (I + A) populations. Figure 4 shows that the impact of the infected on the spatial correlation due to the mechanisms of mobility and social interactions is far greater than that induced by the geographical distribution of the population. Indeed, the spatial correlation in the different maps follows the same trend: in the first phase (when the outbreaks are still limited), the spatial correlation increases; then, the epidemic spreads throughout the map up to the epidemic peak moment when the spatial correlation reaches a minimum. Once the plateau phase of the infected curve is reached (day 29), lockdown policies imply that in areas of lower incidence, the total number of infected people decreases faster, thus producing a new increase in spatial correlation.

**Figure 4: Evolution of Moran's I spatial correlation coefficient over time for the known infected (I) and the total infected (I+A) populations**

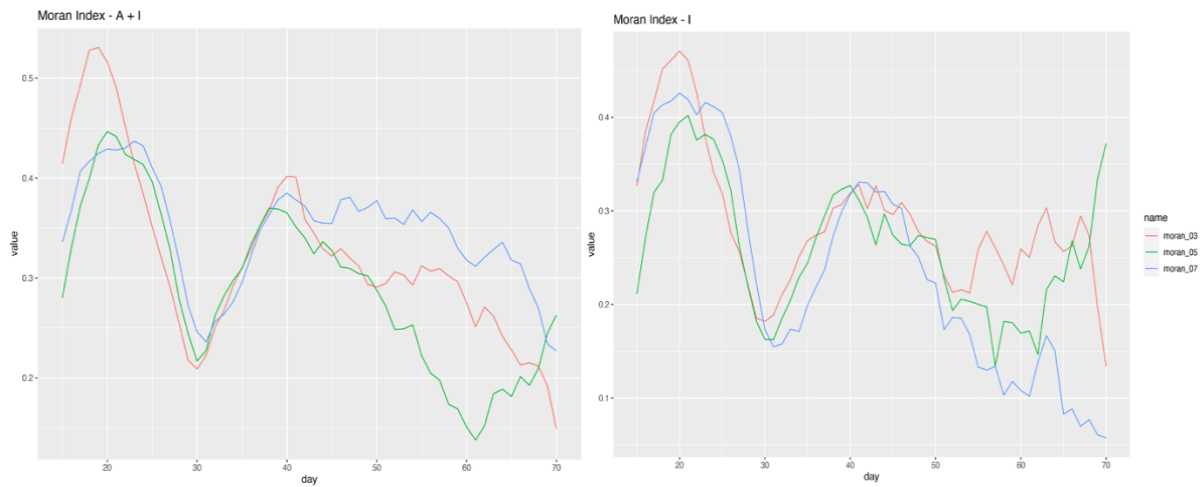

## 5.2 Simulation results

In this section, we present the main results of the sampling simulations. Using the artificial population generated as described in the previous section, we simulate a sample survey in three moments of time, namely, at day 15 (during the ascending phase of the epidemic), at day 29 (at the peak) and at day 43 (during the descending phase in lockdown). For each combination of the parameters of the simulation, we repeated the Monte Carlo simulation for 50,000 runs to ensure the convergence of the sampling methods.

In particular, to measure the speed of the convergence of the different methods, we compare the entropy of each sampling design (Tillé, 2020), defined as follows:

$$I(p) = -\sum_{s \subset U} p(s) \log p(s)$$

where $s$ is a subset of the population, $s \subset U, \mathcal{S} = \{ s \mid s \subset U \}$ represents the set of all possible subsets and $\Pr(S = s) = p(s)$ for all $s \in U$. We assumed that the undetermined point 0 log 0 can be set to 0.

In Table 3, we present the relative entropy attained for each of the methods considered compared with that of the proportional sample design $(I(p)_{lp})$ for only the first stage. We sampled 80 cells in each of the three topical moments (namely, days 15, 29 and 43) and for each map characterized by different levels of spatial correlation ($\rho = 0.3, 0.5,$ and $0.7$) using the five methods: a) the fixed size probability proportional to size (FPPS), b) the local pivotal (LP) method, c) the local cube balanced by the verified infected (LCBV), d) the local cube balanced on the geography of the phenomenon (LCBG) and e) the local cube balanced by the verified infected and on the geography of the phenomenon (LCBVG). The results reported in Table 3 show that for high spatial autocorrelation values, the entropy of the spatial methods is high.

Table 3: Relative entropy of the sampling design.

| Day | $\rho$ | $I(p)_{FPPS}/I(p)_{LP}$ | $I(p)_{FPPS}/I(p)_{LCBV}$ | $I(p)_{FPPS}/I(p)_{LCBG}$ | $I(p)_{FPPS}/I(p)_{LCBVG}$ |
|---|---|---|---|---|---|
| 15 | 0.3 | 0.47 | 0.33 | 0.40 | 0.50 |
| 15 | 0.5 | 1.14 | 0.61 | 1.18 | 1.16 |
| 15 | 0.7 | 1.26 | 0.78 | 1.30 | 1.27 |
| 29 | 0.3 | 0.54 | 0.85 | 0.60 | 0.70 |
| 29 | 0.5 | 0.94 | 0.67 | 0.92 | 1.09 |
| 29 | 0.7 | 1.31 | 0.86 | 1.32 | 1.39 |
| 43 | 0.3 | 0.41 | 0.86 | 0.57 | 0.47 |
| 43 | 0.5 | 0.96 | 0.68 | 0.99 | 1.00 |
| 43 | 0.7 | 1.29 | 0.83 | 1.30 | 1.30 |

Furthermore, we simulated second-stage sampling with a Monte Carlo experiment for different sample cells (20, 40, 80 and 160) with different numbers of people sampled in each cell (1, 3, 5 and 7) on different days (days 15, 29 and 43). For simplicity, in Table 4, we report the results for day 15 (3 people for 80 cells) and display the true value, the estimate obtained as the mean of the simulations, the relative bias expressed in absolute terms (RAB) and the standard error (SE). The results of the simulations obtained using other parameter combinations are not reported because they do not add any further insight and are comparable to those presented here[1]. We can observe that, as expected, all the sampling methods considered produce unbiased estimates. Furthermore, all the methods show consistency, although with very different convergence speeds. All the spatial methods obtained a significantly lower standard error (SE) than proportional sampling. The local cube balanced for the verified infected (LCBV) method displays the lowest error for all spatial autocorrelation values.

---

[1] These results are available from the authors upon request.

Table 4: Simulation results for day 15 (3 people – 80 cells).

| Rho | Sampling method | True value | Estimate | RAB | SE |
|---|---|---|---|---|---|
| 0.3 | FPPS | 1,124 | 1,150 | 0.0222 | 0.36 |
| 0.3 | LP | 1,124 | 1,122 | 0.0021 | 0.32 |
| 0.3 | LCBV | 1,124 | 1,125 | 0.0009 | 0.24 |
| 0.3 | LCBG | 1,124 | 1,123 | 0.0009 | 0.32 |
| 0.3 | LCBVG | 1,124 | 1,125 | 0.0011 | 0.31 |
| 0.5 | FPPS | 1,035 | 1,048 | 0.0125 | 0.37 |
| 0.5 | LP | 1,035 | 1,035 | 0.0001 | 0.34 |
| 0.5 | LCBV | 1,035 | 1,034 | 0.0006 | 0.25 |
| 0.5 | LCBG | 1,035 | 1,037 | 0.0015 | 0.34 |
| 0.5 | LCBVG | 1,035 | 1,028 | 0.0071 | 0.32 |
| 0.7 | FPPS | 1,194 | 1,200 | 0.0052 | 0.34 |
| 0.7 | LP | 1,194 | 1,192 | 0.0020 | 0.30 |
| 0.7 | LCBV | 1,194 | 1,190 | 0.0034 | 0.21 |
| 0.7 | LCBG | 1,194 | 1,188 | 0.0051 | 0.30 |
| 0.7 | LCBVG | 1,194 | 1,193 | 0.0006 | 0.29 |

Complementing the results reported in Table 4, Table 5 compares the standard error (SE) observed in three different epidemic moments: during the ascending phase, at the peak and during the descending phase under the action of the lockdown containment measures. Again, local cube balanced for the verified infected (LCBV) continues to outperform all other methods. These results are in accordance with those obtained in previous simulation studies (Grafström *et al.*, 2013).

Table 5: Simulation results for day 15 (3 people – 80 cells).

| Day | 15 | 29 | 43 | 15 | 29 | 43 | 15 | 29 | 43 |
|---|---|---|---|---|---|---|---|---|---|
| Rho | | 0.3 | | | 0.5 | | | 0.7 | |
| LCBV | 0.24 | 0.55 | 0.38 | 0.25 | 0.58 | 0.42 | 0.21 | 0.48 | 0.36 |
| LCBVG | 0.31 | 0.60 | 0.41 | 0.32 | 0.67 | 0.46 | 0.29 | 0.58 | 0.40 |
| LCBG | 0.32 | 0.63 | 0.42 | 0.34 | 0.68 | 0.47 | 0.30 | 0.60 | 0.42 |
| LP | 0.32 | 0.63 | 0.42 | 0.34 | 0.68 | 0.47 | 0.30 | 0.60 | 0.42 |
| FPPS | 0.36 | 0.67 | 0.45 | 0.37 | 0.72 | 0.50 | 0.34 | 0.65 | 0.44 |

The reason why the local cube balanced for the verified infected (LCBV) method performs the best can be ascribed to the fact that the known infected (I) are distributed in the geographical space similarly to the total infected (A + I), as suggested by the comparison of the Moran index of the two variables shown in Figure 4.

It is important to further remark that the epidemic simulation assumes that the asymptomatic ratio (i.e., the ratio between the known (I) and unknown (A) infected) is constant over space. Although sometimes empirically grounded (Nishura *et al.*, 2020; CDC, 2020; WHO, 2020; World Economic Forum, 2020), this assumption may be too strong in some cases and may not properly represent what happens in reality. Indeed, health screening capacities of areas are often different, which makes data from different health systems less comparable. For this reason, to simulate a heterogeneous screening system in space, we divide the original map into 4 squared macroregions with different abilities to find infected individuals. Heterogeneity can then be induced by assuming different values of the asymptomatic ratio in the four quadrants. In the first (top-right) and third (bottom-left) quadrants, we reduced the number of known infected by 80%, while in the second (top-left) and fourth (bottom-right) quadrants, we reduced the number of unknown infected (A) by 80% and 50%, respectively. In Table 6, we compare the performances of the various sampling methods on day 29 in the map characterized by $\rho = 0.3$.

Table 6: Simulation results for day 15 (3 people – 80 cells).

| Screening | Sampling method | True value | estimate | RAB | SE |
|---|---|---|---|---|---|
| Homogeneous | LCBV | 13,006 | 13,014 | 0.0007 | 0.047 |
| Homogeneous | LCVG | 13,006 | 12,980 | 0.0020 | 0.053 |
| Homogeneous | LCBVG | 13,006 | 13,006 | 0.0000 | 0.052 |
| Heterogeneous | LCBV | 13,006 | 12,982 | 0.0019 | 0.054 |
| Heterogeneous | LCVG | 13,006 | 12,976 | 0.0023 | 0.054 |
| Heterogeneous | LCBVG | 13,006 | 13,041 | 0.0027 | 0.052 |

Adding the heterogeneity, we note that the method that balances for both space and known infected is more robust and performs better than the method that does not consider space and balances only with the number of known infected people. Considering that, in practical instances, the relationship between known and unknown infected cannot be known a priori, a method that also balances for space may be preferable to implement.

## 6. Conclusions and research priorities

Understanding the mechanism of space propagation is fundamental for the evaluation of the spread of the epidemic underway. This is a complex objective that must consider a plurality of aspects. Indeed, the uncertainty of the intensity of the spatial distribution of contagions (and the related risks to the health of the population) depends mainly on the specificity of the syndrome and the fragility of the exposed population. It also depends on the capacity of the health care system and the containment measures introduced and, furthermore, on the behaviour of people in terms of respect for the rules and propensity for mobility. A more or less advanced state of data collection systems represents a further element of uncertainty.
.

The aim of this paper is to improve the current practice in epidemic data collection by introducing a sampling design that exploits the intrinsic peculiarity of data being positively spatially correlated. The main attribute of this method is that it achieves a higher efficiency than both the benchmarking case of a simple random sample and the method of our previous proposal (Alleva *et al.*, 2020).

Thanks to a systematic simulation study, the theoretical properties of the estimators that can be obtained, as proved analytically in the appendices, are confirmed, and the advantages of introducing the spatial dimension appear to be highly significant.

The results obtained encourage us to extend the research in several directions.

Some developments represent a natural extension of the present proposal. In fact, the focus is on the selection of sample units in the first stage. Future developments could also concern the introduction of capture-recapture methods for the second stage. In addition, our simulation studies could be extended by adding more information, such as age, sex, professional condition and other useful characteristics, to balance the sample, thus further improving the efficiency of the estimates. The map structure, population density and mobility schemes could also be introduced to represent different types of urban contexts or regional settlement distributions to tailor the design to different real cases.

Other possible future developments may concern the adaptation of the proposed method for the selection of sample units on which to administer diagnostic tests to trace the diffusion of the virus. One good example is the tracing of the variants of COVID-19 observed in 2020, with a specific focus on their diffusion in a territory.

Finally, the sampling procedure can be adapted to be used as a guide for post-stratification to rebalance the data already collected and obtain more reliable estimates of the phenomenon under study.

**Acknowledgements.** We are very grateful to Giovanni Espa, Paolo Righi and Maria Michela Dickson for their challenging discussions, careful reading of our paper, and useful suggestions, all of which have helped us improve the quality of our proposal.

## Appendix A1: Proof of the variance of the proposed estimator

From Kendall and Stuart (1976, p.196) and Alleva, *et al.* (2021, Appendix 1),

$(A1)$ $E_P E_M (\hat{Y} - Y)^2 = V_P[E_M(\hat{Y}|S)] + E_P[V_M(\hat{Y}|S)] - V_M(Y)$,

where $S$ is the selected sample. We have

$(A2)$ $V_P[E_M(\hat{Y}|S)] = V_P \left[ \sum_{i=1}^{m} \sum_{j=1}^{n_i} \tilde{y}_{ij} w_{ij} \right]$

$= V_{IP} \left( \sum_{i=1}^{m} \tilde{Y}_i w_{Ii} \right) + E_{IP} \left[ V_{IIP} \left( \sum_{i=1}^{m} \hat{\tilde{Y}}_i w_{Ii} \right) \right].$

$V_{IP}(\sum_{i=1}^{m} \tilde{Y}_i w_{Ii})$ is the first-stage sampling variance of the second-stage expectations, and $E_{IP}\left[V_{IIP}\left(\sum_{i=1}^{m} \hat{\tilde{Y}}_i w_{Ii}\right)\right]$ is the first-stage sampling expectation of the second-stage sampling variance, in which

$\tilde{y}_{ij} = m(x_{ij}; \boldsymbol{\beta})$, $\quad \tilde{Y}_i = \sum_{j=1}^{N_i} \tilde{y}_{ij}$, $\quad$ and $\quad \hat{\tilde{Y}}_i = \sum_{j=1}^{\bar{n}} \tilde{y}_{ij} w_{Ii}$.

For variance $V_{IP}(\sum_{i=1}^{m} \tilde{Y}_i w_{Ii})$, we may consider the approximated expression proposed by Deville and Tillé (2005) based on the Poisson sampling theory given by

$(A3)$ $V_{IP}\left(\sum_{i=1}^{m} \tilde{Y}_i w_{Ii}\right) \cong \left[\frac{M}{(M-H)}\right] \sum_{i=1}^{M} \left(\frac{1}{\pi_{Ii}} - 1\right) \eta_i^2$

where

$(A3a)$ $\eta_i = \tilde{Y}_i - \pi_i \boldsymbol{d}'_{Ii} \boldsymbol{\phi}$

and

$\boldsymbol{\phi} = \boldsymbol{\Delta}^{-1} \sum_{i=1}^{M} \pi_{Ii} \left(\frac{1}{\pi_{Ii}} - 1\right) \boldsymbol{d}_{Ii} \tilde{Y}_i$, with $\boldsymbol{\Delta} = \sum_{k \in R} \boldsymbol{d}_{Ii} \boldsymbol{d}'_{Ii} \pi_{Ii}(1 - \pi_{Ii})$.

The above variance resembles the variance expression of the HT estimator under Poisson sampling design conditions, but it uses the residuals, $\eta_i$, instead of the original values, $\tilde{Y}_i$.

Furthermore,

$(A4)$ $E_{IP}\left[V_{IIP}\left(\sum_{i=1}^{m} \hat{\tilde{Y}}_i w_{Ii}\right)\right] = \sum_{i=1}^{M} w_{Ii} \sigma^2_{\tilde{y}_i} N_i \left(\frac{N_i - \bar{n}}{\bar{n}}\right),$

with

$(A5)$ $\sigma^2_{\tilde{y}_i} = \frac{1}{N_i - 1} \sum_{j=1}^{N_i} \left(\tilde{y}_{ij} - \frac{\tilde{Y}_i}{N_i}\right)^2.$

Thus, assuming that $[M/(M-H)] \cong 1$, then, we have

$(A6)$ $V_P[E_M(\hat{Y}|S)] = \sum_{i=1}^{M} \left\{ \frac{1}{\pi_{Ii}} \left[ \eta_i^2 + \sigma^2_{\tilde{y}_i} N_i \left( \frac{N_i - \bar{n}}{\bar{n}} \right) \right] \right\} - \eta_i^2.$

The inner term of the second addendum on the right-hand side of Formula $(A1)$ is

$(A7)$ $V_M(\hat{Y}|S) \cong V_M \left( \sum_{i=1}^{m} \sum_{j=1}^{\bar{n}} y_{ij} w_{ij} \right) = V_M \left( \sum_{i=1}^{M} w_{Ii} \lambda_i \sum_{j=1}^{N_i} y_{ij} w_{IIi} \lambda_{j|i} \right)$

$= \sum_{i=1}^{M} V_M (\hat{Y}_i w_{Ii} \lambda_i) + \sum_{i=1}^{M} \sum_{\ell \neq i}^{M} Cov_M (\hat{Y}_i w_{Ii}, \hat{Y}_\ell w_{I\ell} \lambda_i \lambda_\ell),$

with $\lambda_i = 1$ if $i \in S_I$ and $\lambda_i = 0$; otherwise, $\lambda_{j|i} = 1$ if $j \in S_{IIi}$, and $\lambda_{j|i} = 0$ otherwise. We have

(A8a) $V_M(\hat{Y}_i w_{Ii} \lambda_i) = w_{Ii}^2 \lambda_i \left[ \sum_{i=1}^{N_i} \sigma_u^2 w_{IIi}^2 \lambda_{j|i} + \sum_{j=1}^{N_i} \sum_{k \neq j}^{N_i} \sigma_u^2 \, \rho[d(ij, ik)] w_{IIi}^2 \lambda_{j|i} \lambda_{k|i} \right]$ and

(A8b) $Cov_M(\hat{Y}_i w_{Ii}, \hat{Y}_\ell w_{I\ell} \lambda_i \lambda_\ell) = w_{Ii} w_{I\ell} \lambda_i \lambda_\ell \sum_{j=1}^{N_i} \sum_{k=1}^{N_\ell} \sigma_u^2 \, \rho[d(ij, \ell k)] w_{IIi} w_{II\ell} \lambda_{j|i} \lambda_{k|\ell}$.

For the sampling expectation of $V_M(\hat{Y}_i w_{Ii} \lambda_i)$,

(A9) $E_P[V_M(\hat{Y}_i w_{Ii} \lambda_i)] = \frac{1}{\pi_{Ii}} \left[ \sum_{i=1}^{N_i} \sigma_u^2 w_{IIi} + \sum_{j=1}^{N_i} \sum_{k \neq j}^{N_i} \sigma_u^2 \, \rho[d(ij, ik)] w_{IIi}^2 \pi_{IIij, IIik} \right]$,

where $\pi_{IIij,IIik}$ is the second-stage joint inclusion probability of units $ij$ and $ik$. For simple random sampling without replacement in the second stage,

$\pi_{IIij,IIik} = \frac{\bar{n}}{N_i} \frac{(\bar{n}-1)}{(N_i-1)}$. and

(A9a) $E_P[V_M(\hat{Y}_i w_{Ii} \lambda_i)] = \frac{1}{\pi_{Ii}} \left[ N_i^2 \frac{\sigma_u^2}{\bar{n}} + \sigma_u^2 \frac{N_i}{\bar{n}} \frac{(\bar{n}-1)}{(N_i-1)} \sum_{j=1}^{N_i} \sum_{k \neq j}^{N_i} \rho[d(ij, ik)] \right]$.

The sampling expectation of $Cov_M(\hat{Y}_i w_{Ii}, \hat{Y}_\ell w_{I\ell} \lambda_i \lambda_\ell)$ is

(A10) $E_P[Cov_M(\hat{Y}_i w_{Ii}, \hat{Y}_\ell w_{I\ell})] = \frac{\pi_{Ii,I\ell}}{\pi_{Ii} \pi_{I\ell}} \sum_{j=1}^{N_i} \sum_{k=1}^{N_\ell} \sigma_u^2 \, \rho[d(ij, \ell k)]$.

Since $[(\bar{n}-1)N_i / \bar{n}(N_i - 1)] \cong 1$, the term $E_P[V_M(\hat{Y}|S)]$ is

(A11) $E_P[V_M(\hat{Y}|S)] \cong \sum_{i=1}^{M} \frac{1}{\pi_{Ii}} \left\{ \left[ N_i^2 \frac{\sigma_u^2}{\bar{n}} + \sum_{j=1}^{N_i} \sum_{k \neq j}^{N_i} \sigma_u^2 \rho[d(ij, ik)] \right] + \left[ \sum_{\ell \neq i}^{M} \frac{\pi_{Ii,I\ell}}{\pi_{I\ell}} \sum_{j=1}^{N_i} \sum_{k=1}^{N_\ell} \sigma_u^2 \, \rho[d(ij, \ell k)] \right] \right\}$.

Finally, we have

(A12) $V_M(Y) = \sum_{i=1}^{M} \left\{ \left[ N_i^2 \sigma_u^2 + \sum_{j=1}^{N_i} \sum_{k \neq j}^{N_i} \sigma_u^2 \rho[d(ij, ik)] \right] + \left[ \sum_{\ell \neq i}^{M} \sum_{j=1}^{N_i} \sum_{k=1}^{N_\ell} \sigma_u^2 \, \rho[d(ij, \ell k)] \right] \right\}$.

Considering the expressions $(A6)$, $(A11)$ and $(A12)$, we obtain expression A12.

### Appendix A2 : Efficiency of the proposed design

Expression (13) directly shows that the joint inclusion probabilities $\pi_{Ii,I\ell}$ must be as small as possible when $\rho_{i,\ell}$ is large. This confirms the rule of spatial sampling in which the joint selection of units that are positively correlated should be avoided. As the correlated units are generally geographically close, the sample must be well-spread (or spatially balanced) in the territory. In this way, the term $\sum_{\ell \neq i}^{M} \frac{\pi_{Ii,I\ell}}{\pi_{I\ell}} \sigma_u^2 \rho_{i,\ell}$ vanishes.

At the same time, well spreading the $\bar{n}$ units in second-stage sampling makes component $\rho_{i,i} \sigma_u^2$ of $\sigma_i^2$ very small.

We may obtain a further gain in efficiency by balancing first-stage sampling on the total of the $x_{ij}$ variables. In this way, the balancing equations would be defined on augmented auxiliary vector $\mathbf{d}_{Ii,aug} = (\mathbf{d}'_{Ii}, \mathbf{x}'_{Ii})'$. In such a way, the terms $\eta_i^2$ and $\sigma_i^2$ would be close to 0.

We can obtain an extra gain in the efficiency by balancing second-stage sampling on the $x_{ij}$ **variables** instead of *drawing the units* via simple random sampling without a replacement procedure. The second-stage balancing equations would be

$$(A14) \quad \sum_{j \in S_{IIi}} \frac{x_{ij}}{\pi_{IIi}} \cong x_{Ii}.$$

If we draw the second-stage sample with the cube method, where the balancing equations are defined in Formula (20), the terms $\sigma_{\hat{y}_i}^2$ in formula (13) would be close to 0.

If the above rules of spreading and balancing are followed, we have

$$(A15) \quad E_P E_M (\hat{Y} - Y)^2 \cong \sum_{i=1}^{M} \frac{1}{\pi_{Ii}} N_i^2 \sigma_{eff,i}^2 - F_{eff},$$

where

$$(A16) \quad \sigma_{eff,i}^2 = \frac{\sigma_u^2}{\bar{n}} \text{ and } F_{eff} = \sum_{i=1}^{M} N_i^2 \sigma_u^2.$$

With the constraint that the expected first-stage sample size is fixed, that is,

$$\sum_{i=1}^{M} \pi_{Ii} = m$$

and by using a Lagrangian function, we can determine that $\pi_{Ii}$ values which ensure the minimum of Equation (A1) are

$$(A17) \quad \pi_{Ii} = mN_i \left(\sigma_u/\sqrt{\bar{n}}\right) / \sum_{\ell=1}^{M} N_\ell \left(\sigma_u/\sqrt{\bar{n}}\right) = \frac{mN_i}{N}$$

provided that $mN_i(\sigma_u/\sqrt{\bar{n}}) \leq \sum_{\ell=1}^{M} N_\ell(\sigma_u/\sqrt{\bar{n}})$. Equation A17 defines the conditions under which selecting a PPS first-stage sampling design is efficient.